\documentclass[aps,10pt, pre,reprint,superscriptaddress,showpacs]{revtex4-1}
\usepackage{graphicx}
\usepackage{amsmath}
\usepackage{soul,xcolor}
\usepackage{color}

\begin{document}

%\preprint{}

\title{
%Velocity Correlations and Diffusion in a Strongly Coupled Plasma\\
%or\\
Experimental measurement of non-Markovian dynamics and self-diffusion in a strongly coupled plasma
}


\author{T. S. Strickler}
%\thanks{}
\affiliation{Rice University, Department of Physics and Astronomy, Houston, Texas 77005}
\author{T. K. Langin}
%\thanks{}
\affiliation{Rice University, Department of Physics and Astronomy, Houston, Texas 77005}
\author{P. McQuillen}
%\thanks{}
\affiliation{Rice University, Department of Physics and Astronomy, Houston, Texas 77005}
\author{J. Daligault}
%\thanks{}
\affiliation{Theoretical Division, Los Alamos National Laboratory, Los Alamos, New Mexico 87545, USA}
\author{T. C. Killian}
%\thanks{}
\affiliation{Rice University, Department of Physics and Astronomy, Houston, Texas 77005}

\begin{abstract}
We present a study of the collisional relaxation of ion velocities in a strongly coupled, ultracold neutral
plasma on short timescales  compared to the inverse collision rate.
Non-exponential decay towards equilibrium for the average velocity of a tagged population
of ions heralds
non-Markovian dynamics and a breakdown of assumptions underlying standard kinetic theory.
We prove the equivalence of the average-velocity curve to the velocity autocorrelation
function, a fundamental statistical quantity that provides access to equilibrium transport coefficients and aspects of individual particle trajectories in a regime where experimental measurements have been lacking. From our data, we calculate the ion self-diffusion constant. This demonstrates the utility of ultracold neutral plasmas  for isolating the effects of  strong coupling on collisional processes, which is of interest for dense  laboratory and astrophysical plasmas.

\end{abstract}

\date{\today}

  %52.25.Fi	Transport properties, 52.27.Gr	Strongly-coupled plasmas, 52.72.+v	Laboratory studies of space- and astrophysical-plasma processes
%\keywords{}

\maketitle


\section{Introduction}
In strongly coupled plasmas \cite{ich82}, the Coulomb interaction energy between neighboring particles exceeds the kinetic energy, leading to non-binary collisions that display temporal and spatial correlations between past and future collision events. Such ‘non-Markovian’ dynamics invalidates traditional theory for collision rates \cite{spi62,lan36,bcm12} and calculation of transport coefficients \cite{dal12_mix,dal12_nomix} and frustrates the formulation of a tractable kinetic theory. This challenging fundamental problem is also one of the major limitations to our ability to accurately model equilibration, transport, and equations of state of dense laboratory
 and  astrophysical plasmas \cite{mur04,vho91}, which impacts the design of inertial-confinement-fusion experiments \cite{lab04,dra06}, stellar chronometry based on white dwarf stars \cite{dmu05,gta10}, and models of planet formation \cite{rce05}.
 Molecular dynamics (MD) simulations have been the principal recourse for obtaining a microscopic understanding of short-time collision dynamics in this regime \cite{gma77,hmc75,ham97,rud12,wsh15}, but direct comparison of results with experiment has not been possible.

 In the experiments described here, we  isolate the effects of strong coupling on collisional processes by measuring the velocity autocorrelation function (VAF) for charges in a strongly coupled plasma.
The VAF, a central quantity in the statistical physics of many-body systems, encodes the influence of  correlated collision dynamics and system memory on individual particle trajectories \cite{bzo95}, and it is defined as
\begin{eqnarray}
Z(t)=\frac{1}{3}\langle {\bf v}_k(t)\cdot{\bf v}_k(0)\rangle.
\end{eqnarray}
Here, $\mathbf{v}_k$ is the velocity of particle $k$, and  brackets indicate an equilibrium, canonical-ensemble average.
Remarkably, we obtain this individual-particle quantity from  measurement of the bulk relaxation of the average velocity of a tagged subpopulation of particles in an equilibrium plasma. This contrasts with  measurements of macroscopic-particle VAFs based on statistical sampling of individual trajectories, which is
commonly used in studies of dusty-plasma kinetics \cite{jwt98,nsz06} and Brownian motion \cite{kma73,hct11,ksm14}.

The VAF also provides information on transport processes since its time-integral yields the self-diffusion coefficient through the Green-Kubo relation,
\begin{equation}
\label{eq:KuboGreen}
D=\int_0^{\infty}Z(t)dt\,,
\end{equation}
which describes the long-time mean-square displacement of a given particle through
$D=\lim_{t\to\infty}\langle|{\bf r}(t)-{\bf r}(0)|^2\rangle/6t$ \cite{ein05}. Our results provide the first experimental measurement of the VAF and of self diffusion in a three-dimensional strongly coupled plasma. These results are found to be consistent with MD simulation to within the experimental uncertainty \cite{dal12_mix,dal12_nomix}.

Measurements are performed on ions in an ultracold neutral plasma (UCNP), which is formed by photoionizing a laser-cooled atomic gas \cite{kkb99,kpp07}. Shortly after plasma creation, ions equilibrate in the strongly coupled regime with Coulomb coupling parameter,
\begin{equation}
\label{eq:GammaCouplingPlasma}
 {\Gamma_i}=\frac{e^2}{ 4\pi \varepsilon_0 k_B T}\left(\frac{4\pi n}{3}\right)^{1/3},
\end{equation}
 as large as $\sim 4$. Here, $T$ is the temperature and $n$ is the density.
Electrons in the plasma provide a neutralizing background and static screening on the ionic time scale with a  Debye screening length $\lambda$. This makes UCNPs a nearly ideal realization of a Yukawa one-component plasma \cite{lsm15},  a paradigm model of plasma and statistical physics in which particles interact through a pair-wise Coulomb potential screened by a factor $\mathrm{exp}(-r/\lambda)$.
Ultracold plasmas are quiescent, near local equilibrium, and `clean' in the sense they are composed of a single ion species and free of strong background fields. Strong coupling is obtained at relatively low density, which slows the dynamics and makes short-timescale processes (compared to the inverse collision rate) experimentally accessible.

Powerful diagnostics exist for dense laboratory plasmas. However, their interpretation is complicated by the transient and often non-equilibrium nature of the plasmas, and they do not provide model-independent information on the effects of particle correlations at short timescales. These include, for example,
measurement of the dynamic structure factor with x-ray Thomson-scattering \cite{sgl15,knc08,gln09} and measurement of electrical conductivity using a variety of techniques \cite{sgl15,krs05,asf07,chk13,skj88}. Comprehensive studies of self diffusivity \cite{jwt98,hps09,nsz06,lgo07,obd08} exist for strongly coupled dusty plasmas, but these systems are typically two-dimensional and therefore do not directly illuminate the kinetics of bulk, three-dimensional plasmas.

\section{Methods}

We perform experiments on ultracold neutral plasmas \cite{kpp07}, which are created by first laser-cooling $^{88}$Sr atoms in a magneto-optical trap (MOT) \cite{nsl03}. Atoms are then photoionized with one photon from a narrow-band laser resonant with the principal $^1S_0-^1P_1$ transition at 461 nm and another photon from a tunable 10-ns, pulsed dye laser near 413 nm. The electron temperature ($T_e$) in the plasma is determined by the excess photon energy above the ionization threshold, which can be tuned to set $T_e=1-1000$K. Ions initially have very little kinetic energy, but they possess an excess of Coulomb potential energy, and they equilibrate on a microsecond timescale to a temperature $T_i=0.5-2.5K$, determined primarily by the plasma density \cite{mur03,csl04}.  The ion equilibration process is called disorder-induced heating.

Disorder-induced heating limits the ions to $\Gamma_{i}\approx 2-4$. To obtain measurements on more weakly coupled systems ($\Gamma_i<1$), the plasma is heated with ion acoustic waves \cite{cmk10,kmo12}. Waves are excited by placing a grating (10 cycles/mm) in the path of the ionization beam, which is then imaged onto the MOT for greatest contrast to create a plasma with a striped density modulation. After sufficiently long time, the waves completely damp, heating the plasma and reducing $\Gamma_{i}$.
Electrons provide a uniform screening background for the ions, with screening parameter $\kappa\equiv a/\lambda=0.1-0.55$ in these experiments,
for Wigner-Seitz radius $a=(3/4\pi n)^{1/3}$.

The plasma density distribution is gaussian in shape, $n=n_0\mathrm{exp}(-r^2/2\sigma(t))$, with initial size $\sigma(0)=1$-2 mm. However, due to electron pressure forces, the plasma expands with time dependence given by, $\sigma^2(t) = \sigma^2(0)(1+t^2/\tau^2_{exp})$, where $\tau_{exp}=10 - 50 \,{\mu}s$ is the expansion time \cite{lgs07}.

\begin{figure}
\centering
\setlength\fboxsep{0pt}
\setlength\fboxrule{0.5pt}
\includegraphics[scale=.85]{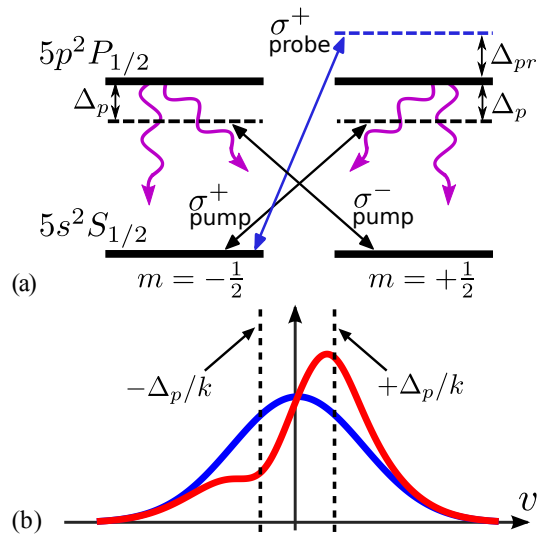}
\caption{(Color) (a) Optical pumping and LIF spectroscopy.  Ions are optically pumped  from the +1/2 to -1/2 electronic spin state around $v_x = -{\Delta_p}/k$, and from the -1/2 to +1/2 spin state around $v_x = +{\Delta_p}/k$ with two counterpropating, circularly polarized laser beams detuned by $-\Delta_p$ rad/s from the $^2S_{1/2}-{^2P_{1/2}}$ transition of the strontium ion (421.7 nm). The velocity profiles of the individual spin populations are  measured with LIF using a tunable circularly-polarized probe beam of variable detuning $\Delta_{pr}$.
(b) Idealized illustration of a pumped velocity distribution for +1/2 ions resulting from optical pumping (red), along with an unperturbed gaussian thermal distribution (blue).}
\label{fig:level}
\end{figure}

An optical pump-probe technique \cite{cbm11,bcm12} is used to measure ${\langle}v_x(t)-\tilde{v}_x(t){\rangle}_+$, the average velocity of a spin-``tagged" subpopulation of ions (labeled $+$) relative to the local bulk velocity of all the ions ($\tilde{v}_x(t)$). The appendix provides a proof that the normalized VAF $\Psi(t)\equiv Z(t)/Z(0)$ is equivalent to the observable ${\langle}v_x(t)-\tilde{v}_x(t){\rangle}_+/{\langle}v_x(0)-\tilde{v}_x(0){\rangle}_+$ as long as the total system is near thermal equilibrium and if terms beyond 2\textsuperscript{nd} order in a Hermite-Gauss expansion of the initial x-velocity distribution function for the subgroup, $f_{x,+}$, are negligible. As shown below, our experiment satisfies these conditions, which provides a new technique for measuring the VAF.

The evolution of ${\langle}v_x(t)-\tilde{v}_x(t){\rangle}_+$  is measured by first using optical pumping to create electron-spin-tagged ion sub-populations with non-zero average velocity (Fig. ${\ref{fig:level}}$). Pumping is accomplished by two counter-propagating, circularly-polarized laser beams each detuned by the same small amount $-\Delta_p$ from the $^2S_{1/2}-{^2P_{1/2}}$ transition at 421.7 nm. Taking advantage of the unpaired electron in the $^2S_{1/2}$ ground state, ions are pumped out of the +1/2 spin state and into the -1/2 state around the negative x-velocity $v_x=-\Delta_p/k$, while ions are pumped from -1/2 to +1/2 spin around the positive velocity $+\Delta_p/k$ (the quantization axis is taken to be along the axis of the pump beams, defined as $\hat{\mathbf{x}}$). This creates subpopulations of +1/2 and -1/2 spin ions having velocity distributions skewed in opposite directions, while the entire plasma itself remains in equilibrium with ${
 \langle}v_x{\rangle}$=0. The pump detuning, $\Delta_p=30$ MHz, is resonant for ions with $|v_x|=12.6$ m/s, which is on the order of the thermal velocity, $v_{th}=\sqrt{k_BT_i/M_i}=13.7$ m/s for $T_i=2$\,K.

We probe the ion distribution with spatially-resolved laser-induced fluorescence (LIF) spectroscopy \cite{lgs07,cgk08} (Fig. ${\ref{fig:level}}$). A LIF probe beam tuned near the $^2S_{1/2}-^2P_{1/2}$ transition of the $^{88}$Sr ion propagates along the x-axis and excites fluorescence that is imaged onto an intensified CCD camera with 1x magnification (12.5 ${\mu}\mathrm{m}$/per pixel), from which the plasma density and x-velocity distribution are extracted.
By using a circularly-polarized LIF probe beam, propagating nearly along the pump beam axis, we selectively probe only the +1/2 ions.

Pumping is applied several plasma periods ($2\pi/\omega_{p}$) after ionization to allow the plasma to approach equilibrium after the disorder-induced heating phase \cite{mur03,kpp07}. ($\omega_{p}=\sqrt{ne^2/\epsilon_0 M_i}\sim 10^7$\,s$^{-1}$ is the ion plasma oscillation frequency.) The optical pumping time is 200 ns, and the pump intensity is 200 mW/cm\textsuperscript{2} (saturation parameter $s_0= 3.5$). LIF data is taken at least 35 ns after the turn off of the pump to avoid contamination of the signal with light from decay of atoms promoted to the $^2P_{1/2}$ state during the pumping process. Electro-optic modulator (EOM) pulse-pickers are used to achieve 10 ns time resolution for application of the pump and probe beams. The pumping and imaging transition is not closed, and about 1/15$^{th}$ of the excitations result in an ion decaying to a metastable $^2D_{3/2}$ state that no longer interacts with the lasers. To ensure that larmor precession of the prepared atomic st
 ates does not contaminate the data, a 4.5 gauss magnetic field is applied along the pump-probe beam axis.

The LIF spectra are fit to a convolution of a Lorentzian function $L(\nu)= \gamma_L/[\pi((\gamma_L)^2+2(\nu)^2)]$ of frequency $\nu$ (Hz) with the one-dimensional ion velocity distribution along the laser axis, $f_{x,+}(v_x)$,
\begin{equation}
\label{eq:convolution}
\int_{-\infty}^{+\infty}(1/\lambda)L(\nu-s/\lambda)f_{x,+}(s)ds,
\end{equation}
where $\lambda$ is the laser wavelength, and $\gamma_L$ is the width of Lorentzian spectral broadening. For this system, $\gamma_L = \gamma_l + \gamma_n\sqrt{1+s_0}$, where $\gamma_l$ is the laser linewidth (5.5 MHz) and $\gamma_n$ is the natural linewidth (20.2 MHz). The width is power-broadened by the laser saturation parameter $s_0$. The distribution $f_{x,+}$ is modeled with a Hermite-Gauss expansion,
\begin{eqnarray}
\label{eq:hermexp}
f_{x,+}(v_x)=&\frac{1}{\sigma_v\sqrt{2\pi}}e^{-\frac{(v_x-\tilde{v}_{x})^2}{2\sigma_v^2}}\sum\limits_{n=0}^N\frac{C_n}{\left(2^nn!\sqrt{\pi}\right)^{1/2}}H_{n}(\frac{v_x-\tilde{v}_{x}}{\sigma_v}).
\end{eqnarray}
where $\sigma_v=\sqrt{k_B T_i/M_i}$, and $H_n$ are Hermite polynomials.
For the analysis, $N=5$ was chosen because the amplitudes of orders 4 and higher were consistent with random noise.

The bulk velocity  of the plasma, $\tilde{v}_x$,
 arises because of the plasma expansion. To separate its effect from thermal velocity spread and perturbation due to optical pumping, the plasma is divided into regions that are analyzed independently \cite{cgk08}. Each region is thin in the direction of the LIF beam propagation (7 overlapping regions $0.1\sigma$ wide, spanning $\pm0.2\sigma$).  $T_i$ and $\tilde{v}_x$ for each region are determined from analysis of LIF data from an unpumped plasma, and only the expansion coefficients $C_n$ are fit parameters. The coupling parameter $\Gamma_i$ varies by no more than $7\%$ across the entire region analyzed.

 \begin{figure}
\centering
{\includegraphics{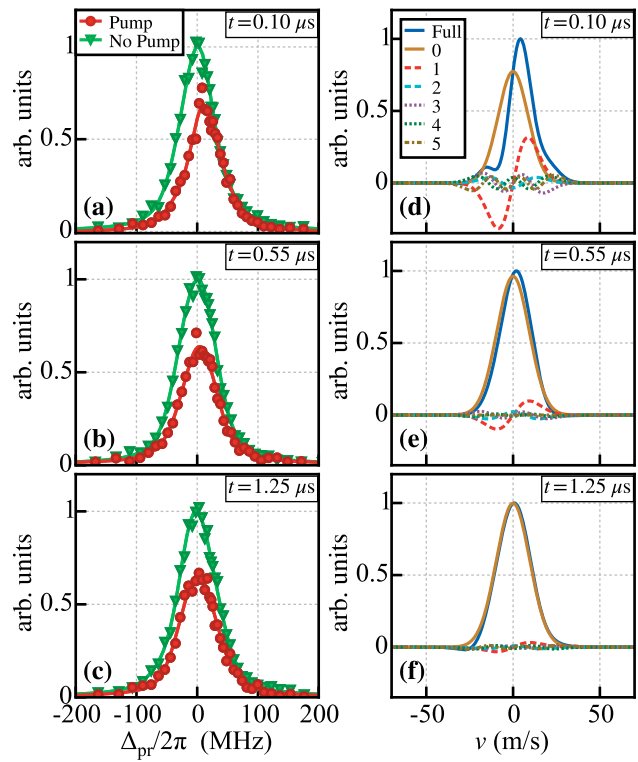}} %\fbox for line around figure
\caption{(Color) (a,b,c) Time evolution of $^2S_{1/2}-^2P_{1/2}$ LIF spectra, pumped (red circles) and unpumped (green squares), for spin +1/2 ions in a plasma with ${\langle}\kappa{\rangle}=0.54$ and ${\langle}\Gamma_{i}{\rangle}=3.3$. Spectra are fit (red and green lines) to a 5th order Hermite-Gauss expansion of the velocity distribution convolved with the lorentzian contribution from the laser and natural linewidths. (d,e,f) Full velocity distributions from the fits of pumped data and constituent Hermite terms n=0-5. Most of the distribution is contained in the first few terms. Terms of order larger than zero decay in time as the distribution relaxes to a Maxwellian.}
\label{fig:sample_spectra}
\end{figure}

\section{Results and Discussion}

Sample LIF spectra at various times after optical pumping are shown in Figs. ${\ref{fig:sample_spectra}}$(a,b,c) for a plasma with ${\langle}\kappa{\rangle}=0.54$ and ${\langle}\Gamma_{i}{\rangle}=3.3$.
Figs. ${\ref{fig:sample_spectra}}$(d,e,f) show the corresponding ion velocity distributions and individual Hermite-Gauss components of the pumped velocity distributions extracted from fits to the raw spectra. At early times, there is significant amplitude in the $n=1$ term, corresponding to the skew in the velocity distribution. This decays away as $f_{x,+}$ approaches a Maxwellian centered around $\tilde{v}_x$. Higher order terms are small at all times, satisfying an important condition for the proof that $\Psi(t)\approx {\langle}v_x(t)-\tilde{v}_x(t){\rangle}_+/{\langle}v_x(0)-\tilde{v}_x(0){\rangle}_+$.

\begin{figure}% \begin{figure*} add asterisk for two-column figure
\setlength\fboxsep{0pt}
\setlength\fboxrule{0.5pt}
{\includegraphics[scale=1]{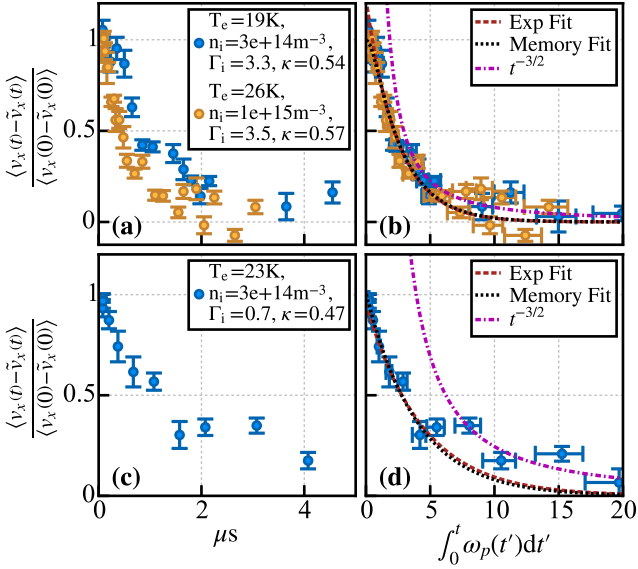}}
\caption{(Color) Relaxation of the average velocity of spin +1/2 ions in an optically pumped plasma. (a) ${\langle}v_x(t)-\tilde{v}_x(t){\rangle}_+/{\langle}v_x(0)-\tilde{v}_x(0){\rangle}_+$ for two data sets with ${\langle}\kappa{\rangle} =0.54$ and ${\langle}\Gamma_{i}{\rangle}=3.3$ plotted versus time, (b). Data from (a) versus time scaled to the time integral of ${\omega}_p$. This plot shows the universal scaling of ${\langle}v_x(t)-\tilde{v}_x(t){\rangle}_+/{\langle}v_x(0)-\tilde{v}_x(0){\rangle}_+$ with ${\omega}_p$ for plasmas of different densities but with approximately the same ${\langle}\Gamma_{i}{\rangle}$ and ${\langle}\kappa{\rangle}$. (c) and (d) are the same plots as (a) and (b) but for ${\langle}\kappa{\rangle} = 0.47$ and ${\langle}\Gamma_{i}{\rangle}=0.7$.
}
\label{fig:scaled_unscaled}
\end{figure}

In the frame co-moving with $\tilde{v}_x$ in a given region, the average velocity of +1/2 ions a time $t$ after cessation of optical pumping is ${\langle}v_x(t)-\tilde{v}_x(t){\rangle}_+=\int f_{x,+}(v_x)(v_x-\tilde{v}_x)dv_x/\int f_{x,+}(v_x)dv_x$. For each plasma, a single time evolution of ${\langle}v_x(t)-\tilde{v}_x(t){\rangle}_+$ is calculated by averaging together individual values of ${\langle}v_x(t)-\tilde{v}_x(t){\rangle}_+$ from each region.
Figs. ${\ref{fig:scaled_unscaled}}$(a,b) show sample data for ${\langle}\kappa{\rangle}=0.54$ and ${\langle}\Gamma_{i}{\rangle}= 3.3$. In Fig. ${\ref{fig:scaled_unscaled}}$(a), data are plotted versus time, while Fig. ${\ref{fig:scaled_unscaled}}$(b) plots ${\langle}v_x(t)-\tilde{v}_x(t){\rangle}_+/{\langle}v_x(0)-\tilde{v}_x(0){\rangle}_+$ versus time scaled by  $\omega_p^{-1}$, showing that this is a universal timescale for the dynamics \cite{lsm15}.
Corresponding data for ${\langle}\kappa{\rangle}= 0.5$, ${\langle}\Gamma{\rangle}=0.7$ are shown in Figs. ${\ref{fig:scaled_unscaled}}$(c,d).

Due to the plasma expansion, the density and thus the plasma frequency $\omega_p$ decreases with time. To account for this, we show the time evolution of the averaged velocity as a function of the scaled time $t_s=\int_0^t \omega_p(t') \mathrm{d}t'$, where the density evolution is described by the self-similar expansion of a gaussian distribution \cite{lgs07}. Similarly, ${\langle}\kappa{\rangle} = \left[{\int_0^{t_{f}}}\kappa(t')\mathrm{d}t'\right]/t_{f}$ and ${\langle}\Gamma_{i}{\rangle} = \left[{\int_0^{t_{f}}}\Gamma(t')\mathrm{d}t'\right]/t_{f}$, where $t_f$ is the total time of measurement.
The density typically varies by a factor of two during the measurement. The normalization factor ${\langle}v_x(0)-\tilde{v}_x(0){\rangle}_+$ is determined from the fit of the data using a memory-function formalism described below.

 %\end{figure*}

\subsection{Observation of Non-Markovian Dynamics}

\begin{figure}
\setlength\fboxsep{5pt}
\setlength\fboxrule{0.5pt}
{\includegraphics[scale=1]{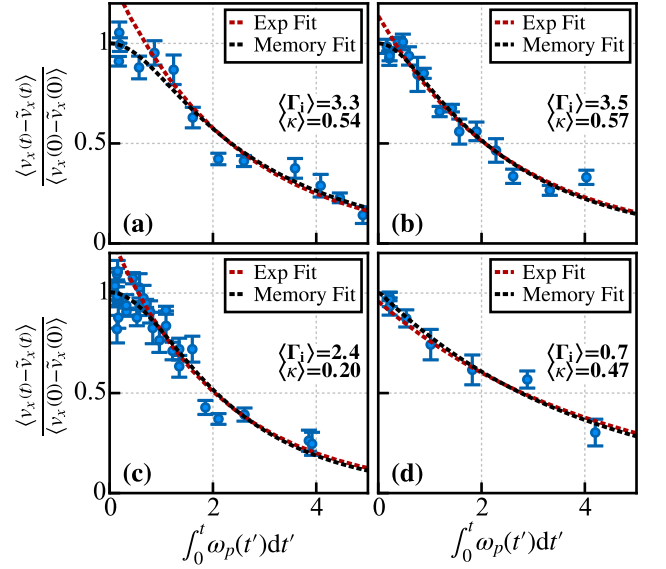}}%\fbox for box around figure
\caption{(Color) Early-time behavior of ${\langle}v_x(t)-\tilde{v}_x(t){\rangle}_+/{\langle}v_x(0)-\tilde{v}_x(0){\rangle}_+$. Lines indicate memory-function and exponential fits. Data and fits for plasmas with (a) ${\langle}\kappa{\rangle}=0.54$ ($T_e = 19K$) and ${\langle}\Gamma_{i}{\rangle}=3.3$, (b) ${\langle}\kappa{\rangle}=0.57$ ($T_e = 26K$) and ${\langle}\Gamma_{i}{\rangle}=3.5$, (c) ${\langle}\kappa{\rangle}=0.2$ ($T_e=105K$) and ${\langle}\Gamma_{i}{\rangle}=2.4$, and (d) ${\langle}\kappa{\rangle}=0.47$ ($T_e=23K$) and ${\langle}\Gamma_{i}{\rangle}=0.7$. Deviation from exponential decay is evident in more strongly coupled plasmas.}
\label{fig:knee}
\end{figure}
Data for ${\langle}v_x(t)-\tilde{v}_x(t){\rangle}_+/{\langle}v_x(0)-\tilde{v}_x(0){\rangle}_+$ shows non-exponential decay of the average velocity up to times  given by $\int_0^t\omega_pdt' \sim 1$,
which is a hallmark of non-Markovian dynamics reflecting the strong coupling of the ions. This is most clearly shown in Fig. ${\ref{fig:knee}}$, which is an expanded view of the early-time data from velocity relaxation curves.
The early-time behavior of $\Psi(t)\approx {\langle}v_x(t)-\tilde{v}_x(t){\rangle}_+/{\langle}v_x(0)-\tilde{v}_x(0){\rangle}_+$ can be described using a memory-function formalism that treats the effects of collisional correlations at the microscopic level \cite{hmc06,bzo95,lve70,hmc75}. It can be derived from a generalized Langevin equation describing the motion of a single test particle experiencing memory effects and fluctuating forces, which is familiar from treatments of Brownian motion\cite{bzo95,hmc06}. The evolution of the VAF is found to be
\begin{equation}
\label{eq:VAF}
\dot{Z}(t)=-\int_0^tK(t-t')Z(t')dt'.
\end{equation}
Here, $K(t-t')$ is the memory function describing the influence at time $t$ from the state of the system at $t'$.

A general, closed-form expression for $K(t-t')$ is lacking, but there are expressions derived from simplifying assumptions that agree well with molecular dynamics simulations for simple fluids \cite{gmi78_1,bzo95}, and Yukawa potentials when $\Gamma>20$ \cite{gma77,gch77}. Some formulas introduce a time constant $\tau$ that may be interpreted as the correlation time for fluctuating forces  \cite{hmc75,hmc06,bzo95}.
If one assumes that collisions are isolated instantaneous events, $\tau\rightarrow0$ and the memory function becomes a delta function. This is the Markovian limit in which the evolution of a system is entirely determined by its present state and $\Psi(t)$ has purely exponential dependence.
Data from a more weakly coupled sample [Fig.\ ${\ref{fig:knee}}$(d), ${\langle}\Gamma_{i}{\rangle}= 0.7$] shows no discernible roll-over in $\Psi(t)$ at short times.

An often-used approximation for $K(t-t')$ for the non-Markovian regime, valid for short times and moderately strong coupling, is the gaussian memory function \cite{hmc75,hmc06,bzo95},
\begin{equation}
\label{eq:memfunction}
K_G(t-t') = \frac{2\gamma_c}{\sqrt{2\pi\tau^2}}\mathrm{exp}\left[-\frac{(t-t')^2}{2\tau^2}\right],
\end{equation}
which satisfies the condition that memory effects vanish at long time and agrees with a Taylor expansion of $K(t-t')$ to second order around $t=t'$ relating $\tau$ to frequency moments of the Fourier transform of the VAF \cite{bzo95}. For the Yukawa OCP, MD simulations have shown that Eq. ${\ref{eq:memfunction}}$ accurately reproduces the ion VAF for $\Gamma<10$ and $\omega_pt < \pi$, and the parameter $\gamma_c$ can be related to a well-defined collision rate \cite{bcm12}.

Figures ${\ref{fig:scaled_unscaled}}$ and ${\ref{fig:knee}}$ show fits of the data in the scaled time range $0<\int_0^t\omega_pdt'<4$ to Eq. ${\ref{eq:VAF}}$ with a gaussian memory function, along with exponential fits of data with $0.8<\int_0^t\omega_pdt'<4.5$. At early times ($\int_0^t\omega_pdt'<0.5$), the memory kernel fit captures the roll-over, which is indicative of non-Markovian collisional dynamics. The values of $\tau$ extracted from the fit are on the order predicted by MD simulations of a classic OCP \cite{hmc75}, although improved experimental accuracy is required before a precise comparison can be made.

\subsection{Ion VAF and Self-Diffusion Coefficients}
\label{DiffSection}
\begin{figure}[t]
\centering
{\includegraphics[scale=1]{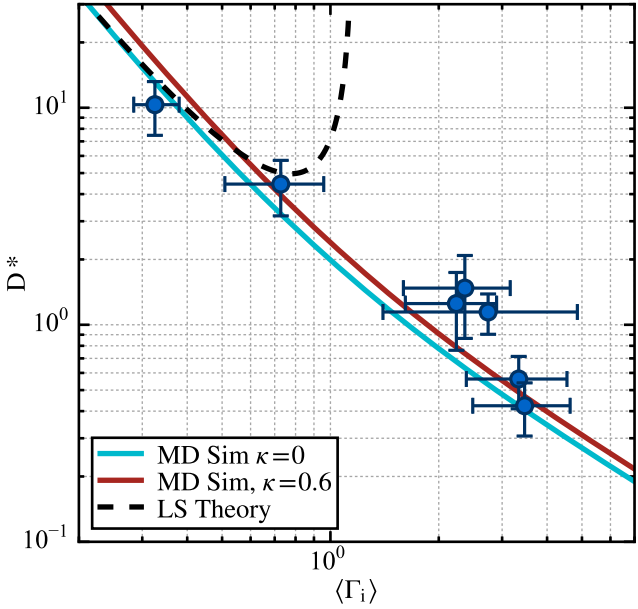}}
\caption{(Color) Plot of the normalized self-diffusion coefficient $D^*$ (Eq. ${\ref{eq:normdiff}}$) calculated for our data, along with upper and lower bounds on the result calculated as explained in Section ${\ref{DiffSection}}$. The solid blue and solid red lines represent results from MD simulations for $\kappa=0$ and $\kappa=0.6$, respectively.\cite{dal12_mix,dal12_nomix}}
\label{fig:diff}
\end{figure}

Using ${\langle}v_x(t)-\tilde{v}_x(t){\rangle}_+/{\langle}v_x(0)-\tilde{v}_x(0){\rangle}_+$ as an approximation for the ion VAF, the self-diffusion coefficient $D$ may be calculated from our measurements.
As is normally the case with calculations of this type, proper treatment of the upper limit of integration in the Green-Kubo formula (Eq.\ ${\ref{eq:KuboGreen}}$) is critical for obtaining accurate results. The behavior of the long-time tail of the VAF for a Yukawa system has not been explored in detail for  the regime of our experiment, and this is an important area for future study. For concreteness, we will assume  a $t^{-3/2}$ dependence  which is  well-established as $t\rightarrow\infty$ for neutral simple fluids \cite{bzo95} and is generally accepted as the slowest possible decay \cite{hrc14}.
We fit the last few data points (beyond the time $t_{cut}$ where $\int_0^{t_{cut}}\omega_p(t')dt'$ $=3\pi$)
to a $bt_s^{-3/2}$ curve, where $b$ is the fit parameter and $t_s=\int_0^{t}\omega_p(t')dt'$ is the scaled time.
The dimensionless, self-diffusion coefficient, $D^*\equiv D/a^2 \omega_p$, is thus calculated as
\begin{eqnarray}\label{eq:normdiff}
D^*\approx\frac{1}{3\left\langle\Gamma_i\right\rangle}\int_0^{t_{s,N}}Z(t_s)dt_s + \frac{2b}{3{\langle}\Gamma_i{\rangle}}t_{s,N}^{-1/2},
\end{eqnarray}
where the first term is calculated numerically from the data by linear interpolation and the trapezoidal rule.
The time of the last data point is $t_N$, and $t_{s,N}=\int_0^{t_{s,N}}\omega_p(t')dt'$ in scaled units.
 Extracted values for $D^*$, along with theoretical curves for $\kappa=0$ and $\kappa=0.6$ determined from a  fit to molecular dynamics simulations   \cite{dal12_mix,dal12_nomix}, are shown in ${\ref{fig:diff}}$.

The contribution from the unmeasured long-time tail of $\Psi(t)$ adds significant uncertainty, which is much greater that our random measurement error.  Here, we take the contribution from the analytic approximation of the tail in Eq.\ \ref{eq:normdiff} as our quoted error bars. The lower error bar thus assumes no contribution beyond our measured points. This is conservative given that
the VAF is exponential in the weakly-coupled limit.
There are additional significant experimental improvements that can be made in the measurement and important systematic effects that must be investigated. The latter are scientifically interesting in their own right, such as the timescale for the approach to equilibrium of velocity correlations after plasma creation and the effect of plasma expansion on the microscopic dynamics. Any complications caused by these systematics can be greatly reduced by performing measurements on a larger plasma for which the expansion time scale is much greater than the characteristic collisional timescale ($\omega_{p} \tau_{exp}\gg 1$).

Other sources of uncertainty include variation of density across the analysis regions, the spread in bulk plasma velocity across and within regions, the time-evolving density, and the uncertainty in the density calibration. These uncertainties are reflected in the horizontal error bars in Fig. ${\ref{fig:diff}}$ and can be significantly reduced in future experiments.

\section{Conclusions}
Utilizing a spin-tagging technique to measure the average velocity ${\langle}v_x(t)){\rangle}_+$ of a subpopulation of ions in an ultracold neutral plasma, and taking advantage of an identification of the time evolution of this quantity with the normalized ion VAF $\Psi(t)$, we have experimentally measured the VAF in a strongly coupled plasma. From this we have calculated the ion self-diffusion coefficient $D$,
which provides an experimental benchmark that has been lacking for molecular dynamics simulations of strongly coupled systems in three dimensions.
The data also display a non-exponential decay of $\Psi(t)$ at early times, which  has not previously been observed experimentally in a bulk plasma and is  indicative of non-Markovian collisional dynamics.
This behavior is well described by a memory-function formalism.

 Overall, these measurements  experimentally validate  foundational concepts  describing how the buildup and decay of ion velocity correlations at the microscopic level determine the dynamics of strongly coupled systems  at the macroscopic level, which cannot be adequately described by simple analytical methods.
 Because ultracold neutral plasmas offer a clean realization of the commonly used Yukawa OCP model, these results are relevant for fundamental kinetic theory and other plasmas  for which effects of strong coupling are important.

\appendix
\section{${\langle}v_x(t)-\tilde{v}_x(t){\rangle}_+/{\langle}v_x(0)-\tilde{v}_x(0){\rangle}_+$ as an approximation to $\Psi(t)$}
\label{app:a}

{
We show that the quantity ${\langle}v_x(t)-\tilde{v}_x(t){\rangle}_+/{\langle}v_x(0)-\tilde{v}_x(0){\rangle}_+$ measured in these experiments corresponds to the normalized VAF $\Psi(t)=Z(t)/Z(0)$ if the initial velocity distribution for $+1/2$ ions ($f_+$) is Maxwellian in $v_y$ and $v_z$ and well-described by a 2\textsuperscript{nd} order Hermite-Gauss expansion in $v_x$, and the optical pumping prepares the subsystem of $+1/2$ ions in a non-equilibrium state close to thermodynamics equilibrium.

To prove this, we transform into the frame co-moving with any bulk hydrodynamic velocity of the ensemble, making $\tilde{v}_x(t)=0$, which does not invalidate any steps in the proof.
For simplicity we assume that the plasma is spatially homogeneous in a volume $V$, although the following arguments can be readily extended to account for non-uniform spatial distributions.
Finally, we assume that the optical pumping occurs instantaneously at some initial time $t=0$.

Let us consider a specified +1/2 ion labeled ``s'' with position ${\bf r}_s(t)$ and velocity ${\bf v}_s(t)$ at time $t$; for a statistical description of the dynamics, it is useful to define the microscopic phase space density is $N_s({\bf r},{\bf v},t)=\delta({\bf r}-{\bf r}_s(t)) \delta({\bf v}-{\bf v}_s(t))$ and its statistical average $f_s({\bf r},{\bf v},t)$.
Before optical pumping, for $t<0$, the system is in thermal equilibrium and $f_s({\bf r},{\bf v},t)=\langle N({\bf r},{\bf v},t)\rangle=f_M({\bf v})/V$, where $f_M({\bf v})$ is the Maxwell-Boltzmann velocity distribution.
After pumping, the \textbf{sub}system is out of thermal equilibrium,
\begin{eqnarray}
f_s({\bf r},{\bf v},t)=f_M({\bf v})/V+\delta\!f_s({\bf r},{\bf v},t).\,
\end{eqnarray}
We assume $|\delta f_s/(f_M/V)|\ll1$, so that $\delta f_s$ satisfies
\begin{eqnarray}
\delta\! f_s({\bf r},{\bf v},t)\!=\!\int{\!d{\bf r}^\prime\!d{\bf v}^\prime\; R_s({\bf r}-{\bf r}^\prime;{\bf v},{\bf v}^\prime;t)\delta\! f_s({\bf r}^\prime,{\bf v}^\prime,0)},\nonumber\\ \label{deltaf}
\end{eqnarray}
where $R_s({\bf r}-{\bf r}^\prime;{\bf v},{\bf v}^\prime;t)$ is the propagator, or retarded Green's function, of the equation that governs the temporal evolution of $\delta f_s$, which is obtained by linearizing the exact evolution equation satisfied by $f_s$.
Remarkably, the propagator $R_s$, which describes the non-equilibrium dynamics of the system, is related simply to the equilibrium time-correlation function $C_s({\bf r}-{\bf r}^\prime;{\bf v},{\bf v}^\prime;t)=\langle N_s({\bf r},{\bf v},t) N_s({\bf r}^\prime,{\bf v}^\prime,t)\rangle$ as follows
\begin{eqnarray}
C_s({\bf r};{\bf v},{\bf v}^\prime;t)&=&\int{\!d{\bf r}^\prime\!\int{\!d{\bf v}^{\prime\prime} R_s({\bf r}-{\bf r}^\prime;{\bf v},{\bf v}^{\prime\prime};t)C_s({\bf r}^\prime;{\bf v}^{\prime\prime},{\bf v}^\prime;0)}}\nonumber\\
&=&R_s({\bf r};{\bf v},{\bf v}^\prime;t)f_M({\bf v}^\prime)/V \label{fluctuationdissipation}\,,
\end{eqnarray}
where, in the last equality, we used the initial value $C_s({\bf r};{\bf v},{\bf v}^\prime;0)=\delta({\bf r})\delta({\bf v}-{\bf v}^\prime)f_M({\bf v})/V$.
This relation is an expression of the fluctuation-dissipation theorem \cite{kro93}.

From Eq.(\ref{deltaf}), the average particle velocity along the x-direction determined in our experiment is
\begin{eqnarray*}
\langle {\rm v}_x(t)\rangle_+&=&\int{\!d{\bf r}d{\bf v}\; {\rm v}_x\,\delta\! f_s({\bf r},{\bf v},t)}\\
&=&\int{\!d{\bf v}\!\int{\!d{\bf v}^\prime\; {\rm v}_x\,\bar{R}_s({\bf v},{\bf v}^\prime;t)\delta\! \bar{f}_s({\bf v}^\prime,0)}}\,,
\end{eqnarray*}
where $\bar{R}_s({\bf v},{\bf v}^\prime;t)=\int{d{\bf r} R_s({\bf r};{\bf v},{\bf v}^\prime;t)}$ and $\delta\!\bar{f}_s({\bf v},0)=\int{\!d{\bf r}\/\delta\!f_s({\bf r},{\bf v},0)}$.
If, as found experimentally (see Fig. \ref{fig:sample_spectra}), $\delta\!\bar{f}_s$ is initially well-described by a 2\textsuperscript{nd} order Hermite-Gauss expansion in $v_x$, then
\begin{eqnarray}
\langle {\rm v}_x(t)\rangle_+&=&\int{\!d{\bf v}\!\int{\!d{\bf v}^\prime\; {\rm v}_x\left(c_0+c_1 {\rm v}_x+c_2{\rm v}_x^2\right)}}\nonumber\\
&&\hspace{2cm}\times\bar{R}_s({\bf v},{\bf v}^\prime;t) f_M({\bf v}^\prime)/V\nonumber\\
&=&\int{\!d{\bf v}\!\int{\!d{\bf v}^\prime\; {\rm v}_x\left(c_0+c_1 {\rm v}_x^\prime+c_2({\rm v}_x^\prime)^2\right)\bar{C}_s({\bf v},{\bf v}^\prime;t)}}\nonumber\,,
\end{eqnarray}
where we used the fluctuation-dissipation theorem (\ref{fluctuationdissipation}) to obtain the second equality.
The zeroth and second order terms vanish since $\bar{C}_s({\bf v},{\bf v}^\prime;t)=\bar{C}_s(-{\bf v},-{\bf v}^\prime;t)$, and the previous result simplifies to
\begin{eqnarray*}
\langle {\rm v}_x(t)\rangle_+&=&c_1\int{\!d{\bf v}\!\int{\!d{\bf v}^\prime\; {\rm v}_x {\rm v}_x^\prime\bar{C}_s({\bf v},{\bf v}^\prime;t)}}\\
&=&c_1\langle v_{x,s}(t) v_{x,s}(0)\rangle=c_1 Z(t)
\end{eqnarray*}
Therefore we obtain the desired result,
\begin{eqnarray*}
\frac{\langle {\rm v}_x(t)\rangle_+}{\langle {\rm v}_x(0)\rangle_+}=\frac{Z(t)}{Z(0)}=\Psi(t)\,.
\end{eqnarray*}
}

\begin{acknowledgments}
This work was supported by the Air Force Office of Scientific Research (FA9550-12-1-0267), Department of Energy,
Fusion Energy Sciences (DE-SC0014455), and Department
of Defense  through the National Defense
Science and Engineering Graduate Fellowship. The work of J. Daligault was supported by the Department of Energy Office of Fusion Energy Sciences.
\end{acknowledgments}

\end{document}